\documentclass[twocolumn,prl]{revtex4-1}
\usepackage[latin9]{inputenc}
\usepackage{color}
\usepackage{textcomp}
\usepackage{amsmath}
\usepackage{amssymb}
\usepackage{graphicx}
\usepackage[unicode=true,pdfusetitle,
 bookmarks=true,bookmarksnumbered=false,bookmarksopen=false,
 breaklinks=false,pdfborder={0 0 0},pdfborderstyle={},backref=false,colorlinks=true]
 {hyperref}
\hypersetup{
 colorlinks,linkcolor=red,citecolor=blue}

\makeatletter

\usepackage{amsfonts}

\makeatother

\begin{document}

\title{Self-locked broadband Raman-electro-optic microcomb }

\author{Shuai Wan$^{1,2,4}$}
\thanks{These authors contributed equally to this work.}
\author{Pi-Yu Wang$^{1,2,4}$}
\thanks{These authors contributed equally to this work.}
\author{Ming Li$^{1,2,4}$}
\thanks{These authors contributed equally to this work.}
\author{Rui Ma$^{3}$}
\author{Rui Niu$^{1,2,4}$}
\author{Fang-Wen Sun$^{1,2,4}$}
\author{Fang Bo$^{3}$}
\email{bofang@nankai.edu.cn}
\author{Guang-Can Guo$^{1,2,4}$}
\author{Chun-Hua Dong$^{1,2,4}$}
\email{chunhua@ustc.edu.cn}

\affiliation{$^{1}$CAS Key Laboratory of Quantum Information, University of Science
and Technology of China, Hefei, Anhui 230026, People's Republic of
China}

\affiliation{$^{2}$CAS Center For Excellence in Quantum Information and Quantum
Physics, University of Science and Technology of China, Hefei, Anhui
230088, People's Republic of China}

\affiliation{$^{3}$MOE Key Laboratory of Weak-Light Nonlinear Photonics, TEDA
Applied Physics Institute and School of Physics, Nankai University,
Tianjin 300457, People's Republic of China}

\affiliation{$^{4}$Hefei National Laboratory, University of Science and Technology
of China, Hefei, Anhui 230088, People's Republic of
China}

\begin{abstract}
Optical frequency combs (OFCs), composed of equally spaced frequency
tones, have spurred advancements in communications, spectroscopy,
precision measurement and fundamental physics research. A prevalent method for generating OFCs involves the electro-optic (EO) effect, i.e., EO
comb, renowned for its rapid tunability via precise microwave field control. Recent advances in integrated lithium niobate (LN) photonics have greatly enhanced the efficiency of EO effect, enabling the
generation of broadband combs with reduced microwave power. However, parasitic nonlinear effects, such as Raman scattering and four-wave mixing, often emerge in high-quality nonlinear devices,
impeding the expansion of comb bandwidth and the minimization of frequency noise.
Here, we tame these nonlinear effects and present a novel type of
OFC, i.e., the self-locked Raman-electro-optic (REO) microcomb by
leveraging the collaboration of EO, Kerr and Raman scattering processes. The
spectral width of the REO microcomb benefits from the Raman gain and
Kerr effect, encompassing nearly $1400$ comb lines spanning over
$300\,\mathrm{nm}$ with a fine repetition rate of $26.03\,\mathrm{GHz}$,
much larger than the pure EO combs. Remarkably, the system
can maintain a self-locked low-noise state in the presence of multiple
nonlinearities without the need for external active feedback. Our
approach points to a direction for improving the performance of microcombs
and paves the way for exploring new nonlinear physics, such as new
laser locking techniques, through the collaboration of inevitable
multiple nonlinear effects in integrated photonics.

\end{abstract}
\maketitle

\section{INTRODUCTION}

Integrated photonics has markedly revolutionized the generation
and manipulation light on a microscale in recent years, heralding
a new era in photonic technologies. One of the most transformative
developments is the realization of optical frequency combs (OFCs)
in on-chip nonlinear microresonators \cite{review2011,review2020,liu2022emerging,1kippenberg2018dissipative,gaeta2019photonic,OFC_application_2}.
OFCs, consisting of evenly spaced discrete spectral lines, have emerged
as essential tools in modern photonics, profoundly transforming fields
like precision timing, spectroscopy, and high-speed communications
\cite{17microwave,16microwave,13communication,14communication,wang2021towards,niu2023khz,tensorM,tensorK}.
Traditional OFCs, generated via mode-locked lasers, require bulky
laboratory setups, limiting their practical applications outside laboratory
environments \cite{modelocked1,modelocked2,modelocked3}. Consequently,
substantial efforts have been made to integrate OFCs into compact,
chip-scale systems \cite{turnkey2020,turnkey2,turnkey3,turnkey4}.

Currently, the primary approaches for generating on-chip OFCs are
based on the Kerr effect, specifically dissipative Kerr solitons (DKSs),
and the electro-optic (EO) effect, known as EO combs. While DKSs enable
ultrafast, passive mode-locking and can facilitate envelope design
and spectrum broadening through dispersion engineering, they require
precise control and are sensitive to thermal fluctuations \cite{1kippenberg2018dissipative}.
In contrast, EO combs, which have been implemented in lithium niobate
resonators and waveguides, are easier to generate and can be directly
locked to microwave frequencies but exhibit limited spectral width
and depend heavily on microwave power \cite{2019_eocomb,zhu2022spectral,LN_x2-2,LN_x2}.
Additionally, the generation of OFCs is often accompanied by parasitic
nonlinear effects, such as Raman scattering \cite{Raman1,Raman2}
in material platforms like lithium niobate (LN) and lithium tantalate
(LT). Raman scattering not only introduces additional frequency noise
but also impedes the generation and extension of OFCs, necessitating
special structural designs to suppress Raman excitations \cite{Ramansoliton,Ramansoliton2,Ramansoliton3,LT}.
To fully leverage the advantages of both DKSs and EO combs while overcoming
their inherent limitations and eliminating the impact of parasitic
nonlinear effects, exploring the synergy of multiple nonlinear effects
is a promising approach. However, the simultaneous occurrence and
synergy of multiple nonlinear effects involve complex physical mechanisms,
making the achievement of high-performance nonlinear OFCs in a single
device a critical unresolved issue.

\begin{figure*}
\centering{}\includegraphics[clip,width=16cm]{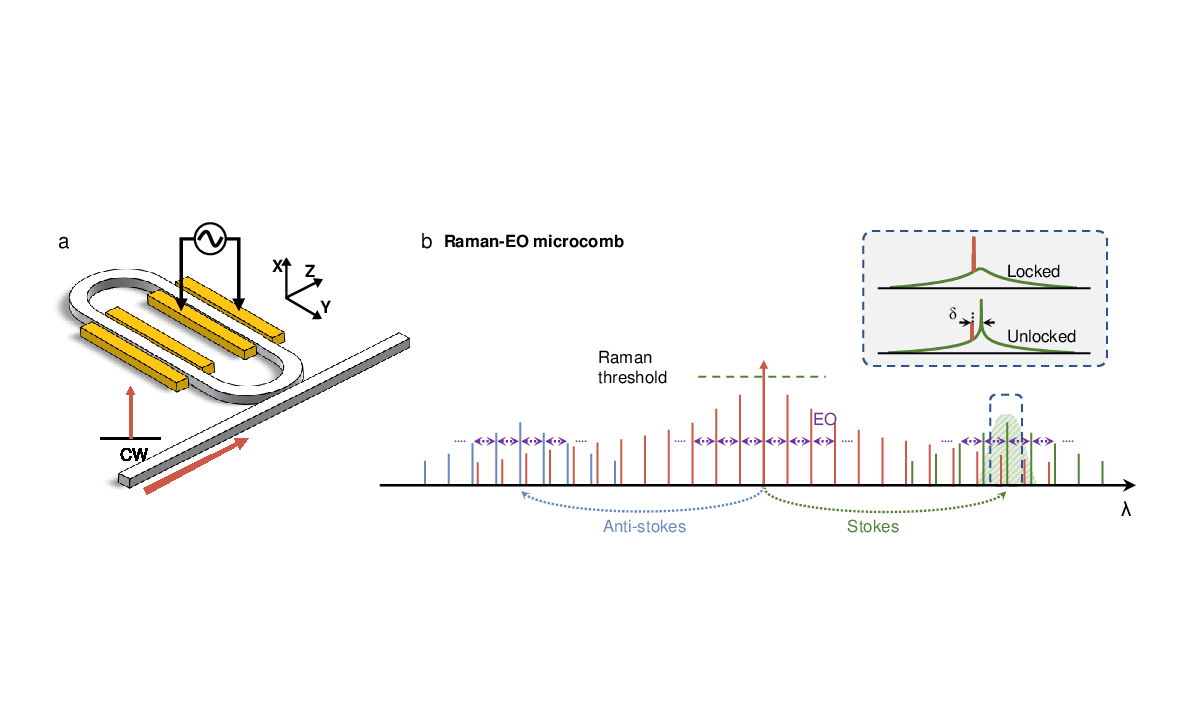}\caption{\label{fig:1} \textbf{Schematic of the hybrid electro-optic comb}.
\textbf{a}. The EO comb is generated by coupling a continuous-wave
(cw) pump laser into a LN resonator under microwave modulation.\textbf{
b}. Spectrum of the EO comb when the pump laser is below the threshold
of Raman lasing.\textbf{ c}. Spectrum of the EO comb when the pump
laser is above the Raman lasing threshold. The EO combs spread around
the pump laser and Raman laser share the same $f_{\text{rep}}$ but
have different $f_{\mathrm{ceo}}$ (unlocked state). With the help
of four wave mixing (FWM), these two combs expand their bandwidths
and overlap. The beating frequency of the two combs in the overlapped
modes is transferred to the Raman laser mode and finally locks the
Raman lasing frequency being integer multiples of the microwave frequency
from the pump frequency (locked state).}
\end{figure*}

Here, we introduce a new type of OFC, by collaborating Raman, EO,
and Kerr effects in a single LN racetrack microresonator to produce
a broadband and flat Raman-electro-optic (REO) microcomb. This novel
mechanism, diverging from systems dependent on single nonlinearities,
employs Raman scattering and EO modulation to generate diverse EO
comb sets, alongside Kerr effect augmentation to broaden the spectral
range. In particular, comb lines originated from the Raman laser,
are self-adaptively phase-locked to the optical and microwave fields
without any external electronic feedback loop, ensuring high coherence
and fast tunability of the whole microcomb. The resulting broadband
REO microcomb, not only showcases an impressive spectral width of
$300\,\mathrm{nm}$, but also maintaining a precise repetition rate
of $26.03\,\mathrm{GHz}$ across nearly $1400$ comb lines. Moreover,
this microcomb features tunable repetition rates and pump laser frequencies,
offering precise control of comb characteristics to enhance performance
and broaden application scopes.

\section{Schematic of self-locked Raman-electro-optic microcomb}

Figure \ref{fig:1} illustrates the configuration of the REO microcomb
generation based on a LN racktrack microresonator. Efficient coherent
energy transfer between adjacent cavity modes occurs when the microwave
modulation frequency aligns with the free spectral range (FSR) of
the microresonator. This transfer, induced by the electro-optic effect,
sequentially shifts the energy of the pump laser to nearby modes one
by one, forming the pure EO comb. While increasing the laser pump
power enhances the power of all comb lines, it does not affect the
EO coupling strength. However, significant Kerr nonlinear processes
can arise and help to expand the spectral width by delivering photons
to different modes via four-wave mixing (FWM). Meanwhile, as the pump
power increases above the Raman threshold, Raman laser can be generated
from solely vacuum noise, which further initiates additional EO combs
under microwave modulation, as shown in Fig. \ref{fig:1}b. However,
the frequency of the Raman lasers depends on the mode detuning and
dissipation, thus the frequency difference with the pump laser may
not be equal to an integer multiple of the microwave frequency. This
leads to a difference ($\delta$) in the carrier-envelope offset frequency
$f_{ceo}$ between the primary comb and the Raman comb, as shown
in the right inset of Fig. \ref{fig:1}b. Moreover, the field intensity
in the pump mode is clamped due to the threshold nature of the Raman
process, which limits the field intensity in all comb lines and eventually
the total power and bandwidth of the EO comb.

Fortunately, the collaboration of these nonlinearities provides a
route to self-adaptively eliminate the $\delta$ and merge these combs
into a single broader microcomb. The mechanism is organized as follows:
(1) Anomalous dispersion in the microresonator enhances FWM, creating
overlapping spectral lines between the primary and Raman combs.
In the overlap region, there are multiple comb lines in a single mode,
with frequencies differed by $\delta$; (2) The Kerr effect transfers
this frequency difference $\delta$ to the Raman gain region, producing
a sideband around the Raman laser. The frequency of this sideband
differs to the pump laser by integer multiples of the microwave frequency;
(3) This sideband competes in the Raman scattering process, potentially
serving as a seed for frequency locking. Proper tuning of microwave
frequency and pump power can reduce $\delta$, leading to the frequency
synchronization of the Raman comb lines with the primary comb
and the alignment of the $f_{ceo}$, as shown in the inset of Fig.
\ref{fig:1}b. This process is analogous to laser injection locking
\cite{turnkey2020} but occurs through coherent nonlinear interactions
within a single device. As a result of the collaboration of these
nonlinearities, the noise from the Raman process is eliminated and
the spectral width of the REO microcomb is substantial broadened.
\begin{figure*}
\centering{}\includegraphics[width=0.95\textwidth]{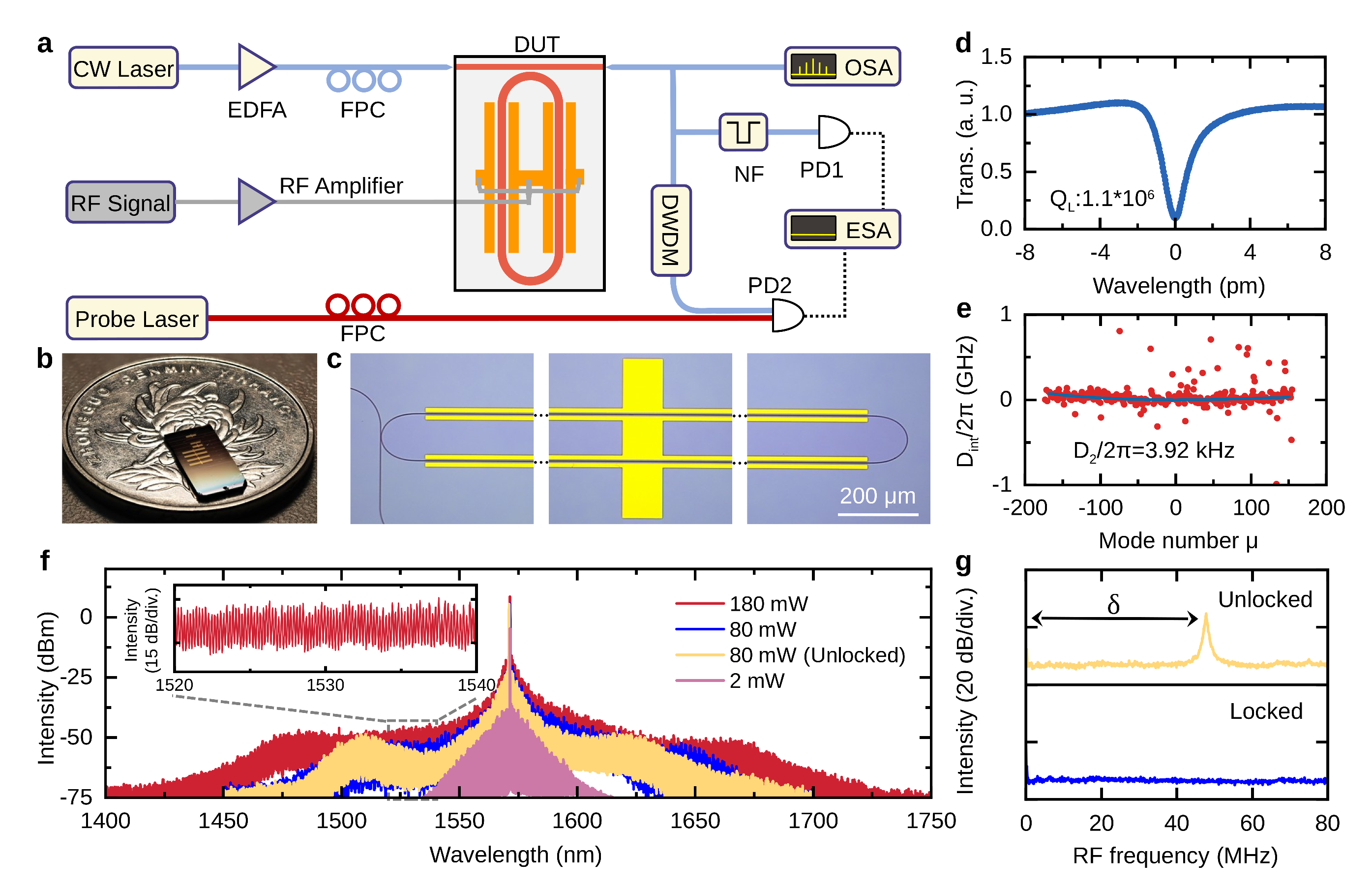}\caption{\label{fig:2} \textbf{a}. Schematic of the experimental setup. EDFA:
erdium doped fiber amplifer; FPC: fiber polarization controller; OSA:
optical spectrum analyzer; DWDM: dense wavelength division multiplexing;
NF: notch filter; ESA: electrical spectrum analyzer. \textbf{b}. Photograph
of the thin film lithium niobate (TFLN) photonic chip. \textbf{c}.
The microscope image of the device consisting of a LN racetrack microresonator
and electrodes. \textbf{d}. The transmission spectrum of a typical
fundamental TE mode with a fitted loaded Q-factors of $1.1\times10^{6}$.
\textbf{e}. Red dots are the dispersion of the microresonator around
$1571\,nm$ measured by the Mach--Zehnder interferometer (MZI).
Blue curve is the fitted curve, with the fitted value of $D_{2}/2\pi$
= $3.92\,\mathrm{kHz}$. \textbf{f}. The optical spectra of the
hybrid microcomb under different on-chip optical pump power. The driven
RF power is $30\,\mathrm{dBm}$. The inset on the upper left corner
shows a flat spectrum between $1520\,$nm and $1540\,$nm. \textbf{g}.
The RF spectra of the output comb field. The disappear of the beating
frequency indicates the whole output field being in a phase-locked
state.}
\end{figure*}

\section{Experimental setup and microcomb characterization}

The experimental setup is shown in Fig. \ref{fig:2}a. The broadband
REO microcomb is experimentally demonstrated using an x-cut thin film
lithium niobate (TFLN) photonic chip, as shown in Fig. \ref{fig:2}b.
Figure \ref{fig:2}c presents the microscope image of the racetrack
microresonator with the FSR of about $26.03\,\mathrm{GHz}$. Detailed
information about the device fabrication can be found in the Methods
section. The pump laser (Toptica CTL 1550) is amplified and polarization-controlled
to match the quasi-TE mode of the on-chip waveguide, and then coupled
into the waveguide supporting the fundamental TE mode through a lensed
fiber. The transmitted light is collected by another lensed fiber
and sent into subsequent characterization instruments. The coupling
loss is approximately $6\,\mathrm{dB}$ per facet. An RF signal, which
matches the FSR of the microresonator, is amplified and applied to
drive the on-chip electrode. Figure \ref{fig:2}d shows the pumped
mode resonates around the wavelength of $1571.04\,\mathrm{nm}$ with
a loaded Q-factor of $1.1\times10^{6}$. The dispersion of the microresonator
around $1571\,\mathrm{nm}$, shown in Fig. \ref{fig:2}e, is analyzed
using the integrated dispersion formula $D_{int}=\frac{1}{2}D_{2}\mu^{2}+\frac{1}{6}D_{3}\mu^{3}+\cdots$,
yielding $D_{2}/2\pi=3.92\,\mathrm{kHz}$ indicating anomalous and
flat dispersion, which is crucial for expanding the spectral width
and locking the microcomb.

\begin{figure*}
\begin{centering}
\includegraphics[clip,width=16cm]{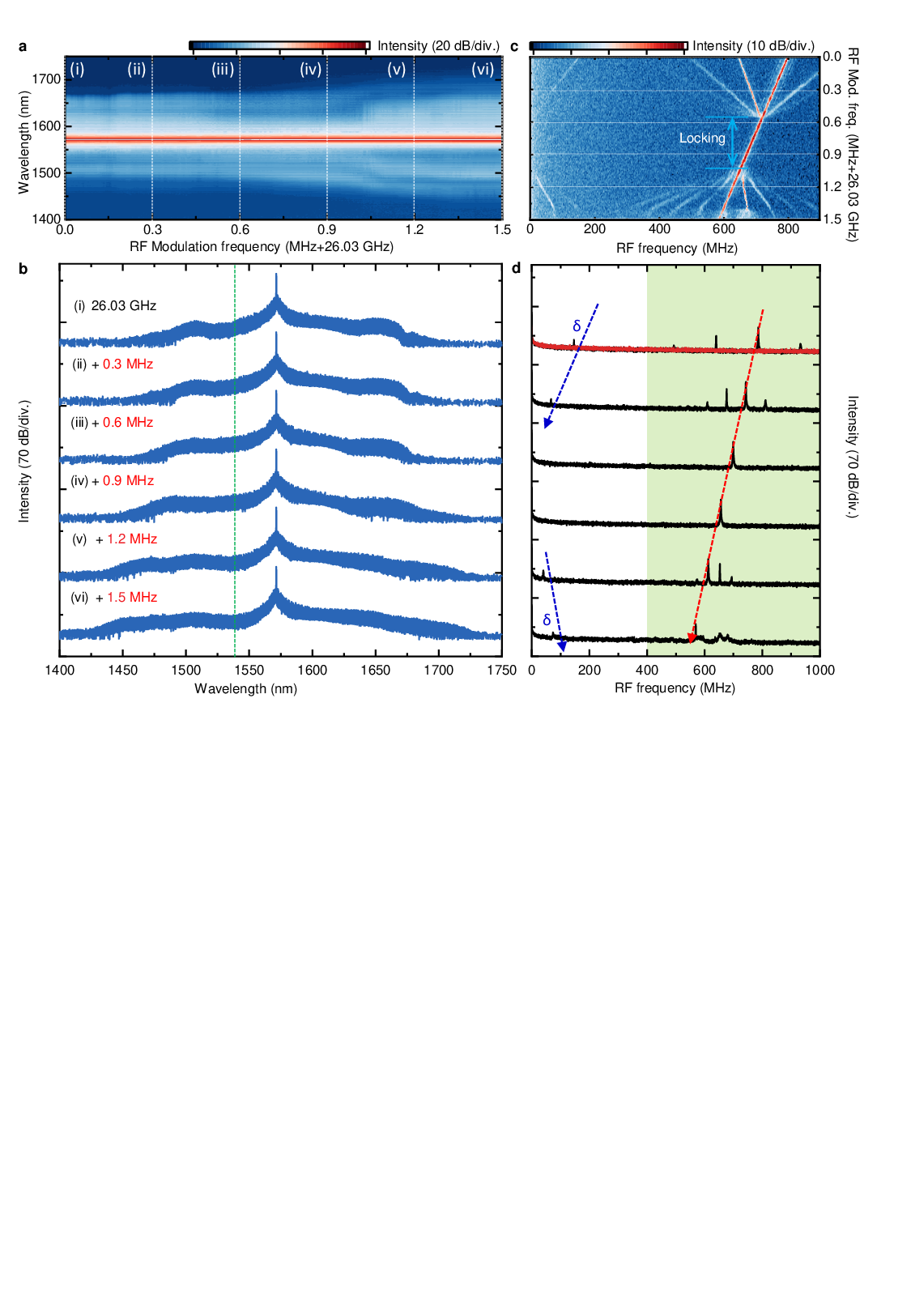}
\par\end{centering}
\centering{}\caption{\label{fig3}\textbf{a},\textbf{b}. The evolution of the optical spectra
when scanning the microwave modulation frequencies. The green line
shows the position of the probe laser.\textbf{ c},\textbf{d}. Measured
beat note signals between comb lines near the green line and a probe
laser at $1541\,\mathrm{nm}$ for different microwave frequencies.\textbf{
}The corresponding beat note signals captured by an ESA, showcasing
three distinct type of spectra, representing self-locked and unlocked
states of the hybrid microcomb.}
\end{figure*}
Figure \ref{fig:2}f displays the optical spectra of microcombs under
different pump conditions. At a low on-chip optical pump power of
$2\,\mathrm{mW}$, the EO effect predominates, producing a typical
EO comb profile with the spectral width of $80\,\mathrm{nm}$ with
the modulation frequency around $26.03\,\mathrm{GHz}$. When the pump power
is increased above the Raman laser threshold, two additional sets
of EO combs, corresponding to the Stokes and anti-Stokes EO combs,
appear on both sides of the primary comb as shown by the yellow
curve in Fig. \ref{fig:2}e. With the help of Raman scattering and
FWM, the spectral width of the hybrid comb becomes significantly wider
compared to the pure EO comb at low pump power. After filtering out
the pump laser, the output comb field is injected to the photodetector
(PD) and analyzed by the electrical spectrum analyzer (ESA). A beat
signal in the upper panel of the Fig. \ref{fig:2}g, indicates a frequency
difference $\delta$ of about $48\,\mathrm{MHz}$ between the primary
comb line and the Raman comb line within the same resonance
mode, implying distinct frequency offsets $f_{\text{ceo}}$ and preventing
treatment as a single phase-locked comb. This beating frequency depends
on the pump and microwave driving conditions as well as environment
noise. Through adjusting the microwave modulation frequency to reduce
the $\delta$, an abrupt disappearance of the beat signal reveals
that these combs is merged into a single one (lower panel of the Fig.
\ref{fig:2}g). The locked REO microcomb shows a broad spectral width
and obvious flatness along the wings, as well as an overall higher
field intensity of all comb lines (blue curve in Fig. \ref{fig:2}f).
With further increaseing optical pump power to $180\,\mathrm{mW}$,
more significant Kerr effect can expand the spectral width further
to over $300\,\mathrm{nm}$.

\section{Locking behavior of the microcomb}

To characterize the self-locked state of the REO microcomb and exclude
any accidental alignment of the primary and Raman combs, we experimentally
investigate the locking dynamics by varying the microwave frequency
under fixed optical pump power ($150\,\mathrm{mW}$) and modulation
power ($30\,\mathrm{dBm}$). The microwave modulation frequency starts
at $26.03\,\mathrm{GHz}$. As the modulation frequency is increased
to better match the FSRs near the pumped mode, the spectral width
of the entire microcomb broadens progressively, as illustrated in
Fig. \ref{fig3}a. Meanwhile, the intensity peak of the comb lines
gradually shifts from longer to shorter wavelengths (from Stokes to
anti-Stokes comb). This shift is attributed to the anomalous dispersion
of the microresonator, where the FSR increases from longer to shorter
wavelengths. Figure \ref{fig3}b provides  a detailed view of how
the spectral intensity distribution varies with the modulation frequency.

Limited by the resolution of the OSA, the details around the overlap
region are challenging to discern. Therefore, a probe laser around
$1541\,\mathrm{nm}$ is employed to detect the comb lines in the overlap
region. The probed comb lines near $1541\,\mathrm{nm}$ are extracted
by filtering other comb lines using dense wavelength division multiplexing
(DWDM). The corresponding beat signals are displayed in Fig. \ref{fig3}c,
d. There are two distinct groups of beat signals in the RF spectra.
The lower one, corresponding to the frequency difference $\delta$
between comb lines, gradually decreases and disappears with increasing
the modulation frequency, and then reappears after hundreds of kHz,
as indicated by the blue dashed line in Fig. \ref{fig3}d. The disappeared
region corresponds to the locking range, which can also be confirmed
by the beat signal of the probe laser with the comb lines in the nearest
cavity mode, i.e., another group of beat signals at higher frequency.
The central peak of this group of beat signals, marked in red dashed
line in Fig. \ref{fig3}d, represents the comb line from the primary
comb, and the slope of this red dashed line represents the mode index
of the probed mode relative to the pumped mode. The peak spacing equals
to the offset difference $\delta$, which scales linearly with the
microwave frequency. When the modulation frequency is outside the
locking range, both the primary and Raman comb lines coexist within
the cavity mode, along with their FWM generated sidebands. Otherwise,
there is only one single comb line within the cavity mode. The sudden
shift of the beat signals and its dependence on the microwave frequency
are clear indicators of the self-locking process. The width of the
corresponding locking range can reach $\mathrm{520\,kHz}$, indicating
the robustness of the self-locked microcomb against small variations
in the pump and microwave fields as well as environment fluctuations.
The self-locked microcomb exhibits a flat distribution of comb lines
at its wings, which is attributed to the effect of Raman gain and
Kerr effect. Furthermore, the self-locking mechanism also reduces
Raman noise (see Supplementary Information), as widely seen in injection
locking experiments \cite{turnkey2020}. In this way, the self-locking
mechanism transforms the typically harmful Raman scattering into a
beneficial nonlinear gain mechanism for broadband EO comb generation.

\begin{figure}
\begin{centering}
\includegraphics[clip,width=8cm]{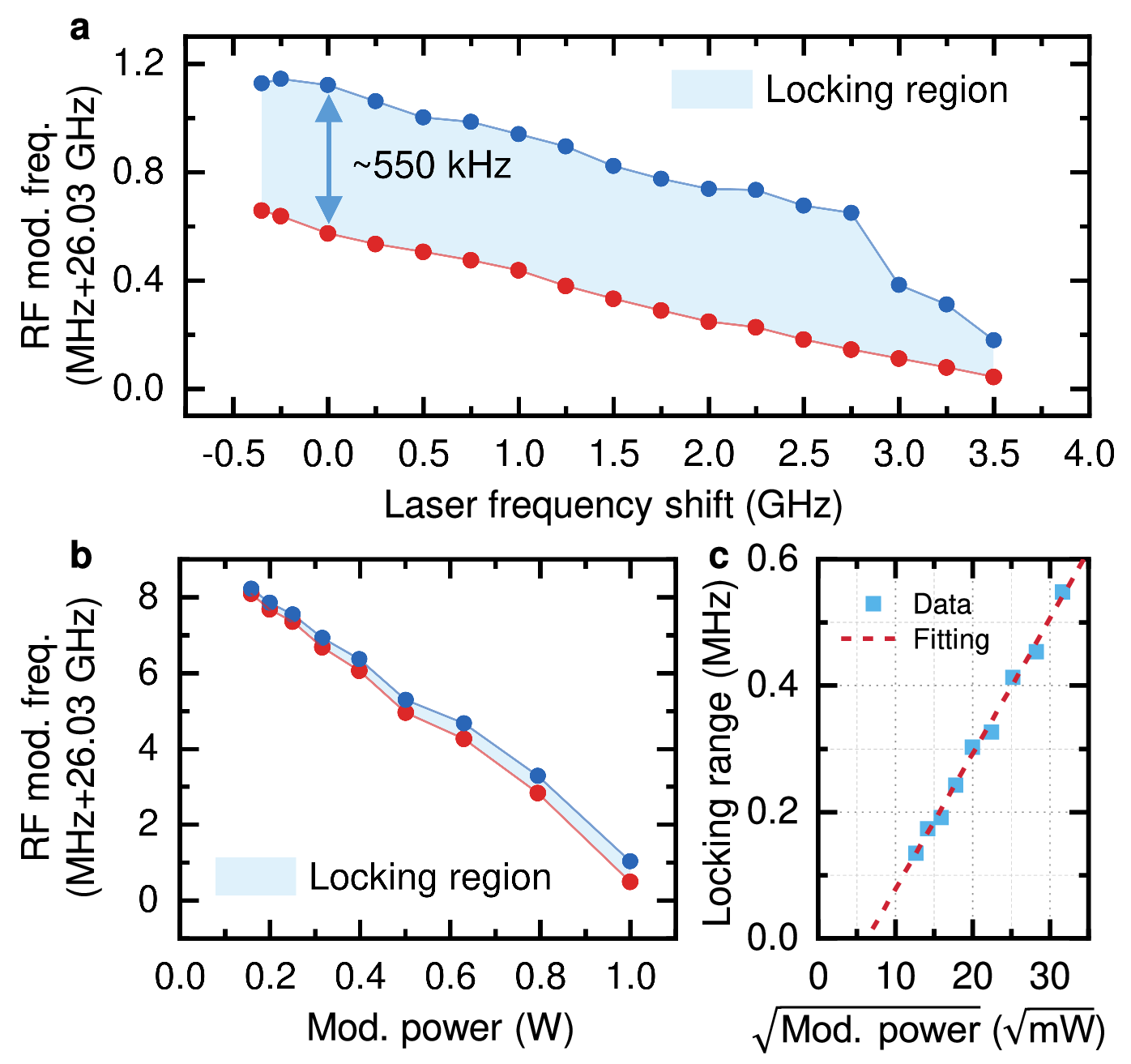}
\par\end{centering}
\centering{}\caption{\label{fig4} \textbf{a}. Measured beat note signal by a probe laser
at $1541\,\mathrm{nm}$ for different pump laser frequencies. The
comb is self-locked in the shaded region between the two liens The
locking range shift with laser frequency shift, demonstrating the
self-locked comb works for various $f_{\text{rep}}$. b. Relationship
between the two boundaries of the locking range and the modulatioon
power. \textbf{c}Relationship between the microwave amplitude and
the locking range, which is well fitted by a linear function.}
\end{figure}

Indeed, the locking range of the microcomb offers a precise tuning
method for fine-tuning $f_{\mathrm{rep}}$ across several hundred
kHz. To realize extensive tuning of $f_{\text{rep}}$, it is necessary
to demonstrate that the microcomb can remain locked under various
microwave frequencies that are significantly different. By adjusting
the pump laser frequency, the locking range can be changed, as depicted
in Fig. \ref{fig4}a. It is observed that the locking range gradually
shifts to lower frequency as the pump laser frequency is decreased,
due to the diminishing frequency difference between the pump laser
and the Raman laser. When the pump laser frequency is excessively
high, the detuning between the pump laser and the cavity mode becomes
too large, resulting in insufficient intracavity power to transfer
the offset difference $\delta$ to the Raman laser mode, thus reducing
the locking region. Conversely, when the pump laser frequency is too
low, the pump laser shifts to the red-detuned side of the cavity mode,
causing it to exit the cavity mode. The upper and lower boundaries
at different laser frequencies define a locking region, within which
the locking range exceeds $500\,\mathrm{kHz}$ for most laser frequencies,
with the largest range reaching $550\,\mathrm{kHz}$.

In addition to the optical tuning method, the microwave energy can
heat the chip, thus changing the FSR of the microresonator via the
thermal effect. This alteration significantly impacts the locking
behavior of the microcomb. Figure \ref{fig4}b illustrates how the
two boundaries of the locking range shift when the microwave modulation
power is adjusted from $21\,\mathrm{dBm}$ to $30\,\mathrm{dBm}$,
showing that the locking frequency can vary within a range of $8\,\mathrm{MHz}$.
From the data in Fig. \ref{fig4}b, we also analyzed the relationship
between the width of the locking range and the modulation power. Since
the comb line amplitude is proportional to microwave amplitude of
the EO modulation, a larger microwave amplitude leads to an extended
locking range. As depicted in the Fig. \ref{fig4}c, the locking range
exhibits a positive relationship with the microwave amplitude, fitting
well to a linear function.

\section{Conclusion}

In conclusion, we have demonstrated the generation of a broadband
REO microcomb in a single LN microresonator, achieving a spectral
width exceeding $300\,\mathrm{nm}$ and more than $1400$ comb lines
at a $26.03\,\mathrm{GHz}$ repetition rate. By utilizing the cooperation between
Kerr, Raman and EO nonlinearities , we have addressed the spectral
width and noise limitation imposed by Raman scattering, transforming
it into a beneficial nonlinear process that enhances the width and
the flatness of the optical spectrum. The resulting microcomb is not
only broadband but also highly tunable and capable of rapid on-chip
modulation with a $f_{\mathrm{rep}}$ tuning ranging of $8\,\mathrm{MHz}$.
These characteristics make it an indispensable tool in applications
ranging from high-resolution spectroscopy to advanced telecommunications
and quantum information processing. To further expand the spectral
width and tunability of this microcomb, enhancing the self-locking
capability by increasing the pump conversion efficiency \cite{2022_eocomb},
engineering dispersive waves at the overlap region, and utilizing
higher-order Raman processes are recommended. The self-locking mechanism
can also be adapted for other parametric and competitive nonlinear
processes, such as $\chi^{(2)}$ and $\chi^{(3)}$ parametric oscillation,
Brillioun scattering and lasing process. Given the ubiquity of different
nonlinearities in integrated nonlinear photonics, our findings may
inspire further research into exploiting these interactions to develop
novel nonlinear photonic devices.

\vbox{}

\noindent\textbf{Methods}\\ \textbf{Device fabrication}

The device is fabricated on a commercial x-cut thin film lithium niobate
(TFLN) wafer (NANOLN). The 600-nm-thick TFLN is bonded to a silicon
substrate with 500-\textmu m-thick silicon and 2-\textmu m-thick wet
oxidation silicon dioxide ($SiO_{2}$). Electron-beam lithography
is used to define the pattern of the device with hydrogen-silsesquioxane
(HSQ) resist.Then, the film is partially etched by argon-ion-based
reactive ion etching in an inductively coupled plasma (ICP) etcher
to form a 350-nm-depth trapezoidal waveguide cross section with a
remaining slab of $250\,\mathrm{nm}$. The whole chip is cleaned by
buffered HF solution and RCA1 cleaning solution (NH3:H2O2:H2O = 1:1:5)
to remove the remaining HSQ resist and the redeposition formed in
the etching process. The gold modulation electrodes are patterned
using laser direct writing and the metal (15 nm of chromium, 300 nm
of gold) is transferred using thermal evaporation and the bilayer
lift-off process. The LN racetrack microresonator used in experiments
has a top width of $1.4\,\mathrm{\mu m}$and a sidewall angle of around
$60\circ$. The fiber-to-chip coupling loss is approximately $\mathrm{6\,dB}$
per facet.

\vbox{}

\noindent\textbf{Data availability}\\ All data generated or analyzed
during this study are available within the paper. Further source data
will be made available on request.


%

\vbox{}

\noindent\textbf{Acknowledgments}\\ The authors thank C.-L. Zou
for helpful discussions. The work was supported by the National Natural
Science Foundation of China (12293052, 11934012, 12104442, 92050109,
12374361, and 92250302), Innovation program for Quantum Science and
Technology (2021ZD0303203), the CAS Project for Young Scientists in
Basic Research (YSBR-069), the Fundamental Research Funds for the
Central Universities and the USTC Research Funds of the Double First-Class
Initiative. This work was partially carried out at the USTC Center
for Micro and Nanoscale Research and Fabrication.

\vbox{}

\noindent\textbf{Author contributions}\\ \textcolor{black}{S. W.,
P.-Y. W. and M. L. contribute equally to this work. C.-H. D. and S.
W. conceived the experiments, S. W., P.-Y. W., R. M., and F. B. prepared
devices, S. W., P.-Y. W. and R. N. built the experimental setup and
carried out measurements, with assistance from F.W.S. M.L. provided
theoretical supports. S. W., M. L. and C.-H.D. wrote the manuscript
with input from all co-authors. C.-H.D. and G.-C.G. supervised the
project. All authors contributed extensively to the work presented
in this paper}\textcolor{black}{\emph{.}}

\vbox{}

\noindent\textbf{Competing financial interests}\\The authors declare
no competing financial interests.

\end{document}